\documentstyle[12pt]{article}
\def\tr{\mathop{\rm Tr}}
\def\where{\mathop{\rm where}}
\def\and{\mathop{\rm and}}

\begin{document}

\date{ }
\vspace{4cm}
\title{Temperature Derivative of the Superfluid Density in the
Attractive Hubbard model. }

\author{ F.F. Assaad$^1$, W. Hanke$^1$ and D.J. Scalapino$^2$ \\
      $^1$ Physikalisches Institut, Universit\"at W\"urzburg \\
           8700 W\"urzburg, FRG. \\
      $^2$ Department of Physics, University of California \\
           Santa Barbara, CA 93106-9530 }

\maketitle
\begin{abstract}
Based on extensions of the grand-canonical Quantum Monte-Carlo algorithm
to incorporate magnetic fields,
we provide numerical data confirming the existence of a Kosterlitz-Thouless
transition in the attractive Hubbard model.
Here,
we calculate the temperature derivative of the superfluid density,
$ \frac{ \partial \beta D_s(\beta) } {\partial \beta} $,  to pin down
the transition.
The latter quantity is obtained from the difference in internal energy
of systems which differ by a phase twist $\pi/2$ in the boundary condition
along one lattice direction.
Away from half-band filling, $ \frac{ \partial \beta
D_s(\beta) } {\partial \beta}$  shows a response which increases with
lattice size at the
transition temperature. In contrast, such a signal is not observed
for the case of a  half-band filling.

\end{abstract}

PACS numbers: 75.10.Jm

\newpage

Numerical simulations of the attractive Hubbard Hamiltonian
\cite{Scalettar,Moreo,Moreo1}, indicate
that this fermionic lattice model has an s-wave superconducting ground state
in two-dimensions.
A finite-size scaling analysis (up to one hundred sites) of the
s-wave pair-pair
correlation functions has provided some evidence that the model
has a finite-temperature Kosterlitz-Thouless (KT) \cite{KT}  transition to
a superconducting state away from half-band filling \cite{Moreo1}.
Based on extensions of the grand-canonical Quantum
Monte-Carlo (QMC) algorithm,
we present numerical data which clearly confirm the
conjecture of a KT transition.
Our approach relies on the calculation of the temperature derivative
of the superfluid density, $ \frac{ \partial \beta
D_s(\beta) } {\partial \beta}$, where $ \beta $ is the inverse
temperature.  In the framework of such a
KT transition, $D_s(\beta)$  shows a universal jump at
the transition temperature \cite{KT1}. Hence, $ \frac{ \partial \beta
D_s(\beta) } {\partial \beta}$ in the thermodynamic limit
behaves like a Dirac delta-function at the transition temperature,
and even on finite-sized clusters is expected to yield a very clear
signal. Compared to previous estimations of the $T=0$ superfluid density ,
via flux quantization \cite{Assaad} or linear response \cite{Scalapino},
the method described here leads to  numerical results which exhibit
a peak at $T_{KT}$ which grows as  the size of the system increases.
This provides a clear signal of the KT transition.
Such signatures of the KT transition have already been observed in
classical \cite{Himbergen} as well as in quantum XY models
\cite{Loh}.
In this letter it is shown that, for the attractive Hubbard model,
$ \frac{ \partial \beta D_s(\beta) }
{\partial \beta}$
is given by the difference in internal energy  of systems
which differ by a phase twist $\pi/2$ in the boundary condition
along one lattice direction.
The internal energies are extracted from  finite temperature
quantum Monte-Carlo (QMC)  data on cluster sizes ranging
from $4 \times 4$ to $ 8 \times 8$.
Comparison of the data for different lattice sizes,
strongly supports the existence
of a KT transition away from half band filling.

The attractive Hubbard hamiltonian we consider is given by:
\begin{eqnarray}
\label{Ham1}
     H(\Phi)  = & - & t \sum_{\vec{i},\sigma}
   \left( \exp\left( \frac{2\pi i \Phi}{L\Phi_{0}} \right)
        c_{\vec{i},\sigma}^{\dagger} c_{\vec{i} + \vec{a}_{x}, \sigma } +
        \exp\left( -\frac{2\pi i \Phi}{L\Phi_{0}} \right)
        c_{\vec{i} + \vec{a}_{x},\sigma}^{\dagger} c_{\vec{i}, \sigma }
  \right) \nonumber \\
          & - &   t \sum_{\vec{i},\sigma}
   \left( c_{\vec{i},\sigma}^{\dagger} c_{\vec{i} + \vec{a}_{y},\sigma} \;+\;
          c_{\vec{i} + \vec{a}_{y},\sigma}^{\dagger} c_{\vec{i},\sigma }
   \right) \nonumber \\
         & + & U \sum_i \left( n_{i,\uparrow} - \frac{1}{2} \right)
                      \left( n_{i,\downarrow} - \frac{1}{2} \right)
          -\mu \sum_i \left( n_{i,\uparrow} + n_{i,\downarrow} \right),
\; \; \;  U < 0.
\end{eqnarray}
Here,  $c_{\vec{i},\sigma}^{\dagger}$  creates an electron with
z-component of spin $\sigma$ on lattice site $\vec{i}$,
$\vec{a}_{x,y} $ are the lattice vectors of unit length, $\Phi_0 =
\frac{hc}{e}$ is
the flux quantum, and $L$ is
the linear length of the square lattice. The fermionic operators
obey periodic boundary conditions in both lattice directions.
The above Hamiltonian thus describes an attractive Hubbard model on a
torus.
The phase factors in the kinetic energy term
account for the presence of a magnetic field of flux $\Phi$
threaded through the center of the torus. Through a canonical
transformation,  one may remove the  phase factors in (\ref{Ham1}).
However, under this canonical transformation, the
x-boundary acquires a phase twist  $ \exp \left( 2 \pi i \Phi / \Phi_0
\right) $ \cite{Byers}.  Hence, the threaded magnetic field acts merely
as boundary effect.

Let us  now consider the free energy,
\begin{equation}
      F(\Phi) = -\frac{1}{\beta} \ln \tr e^{-\beta  H(\Phi)},
\end{equation}
and expand it up to second order in $\Phi/\Phi_0$ around $\Phi=0$. This yields:
\begin{eqnarray}
\label{Expan}
   F(\Phi)&  = &  F(\Phi = 0)  + \left(\frac{ \Phi }{\Phi_0}\right)^2
                                 D_s(\beta) +
       O\left(\left(\frac{ \Phi }{\Phi_0}\right)^4\right)
           \; \; \where \nonumber \\
   D_s(\beta) & = & -   \left( \frac{2\pi}{L} \right)^{2}
         \left( <K_{x}> +
         \int_{0}^{\beta} d\tau <J_{x}(\tau)J_{x}(0)> \right),
         \; \; \nonumber \\
   J_{x} & = & -it \sum_{\vec{i},\sigma}
 \left( c_{\vec{i},\sigma}^{\dagger} c_{\vec{i} + \vec{a}_{x}, \sigma } \; - \;
         c_{\vec{i} + \vec{a}_{x},\sigma}^{\dagger} c_{\vec{i}, \sigma}
         \right) \; \; \and \nonumber \\
   K_{x} & = & -t \sum_{\vec{i},\sigma}
\left( c_{\vec{i},\sigma}^{\dagger} c_{\vec{i} + \vec{a}_{x}, \sigma } \; + \;
       c_{\vec{i} + \vec{a}_{x},\sigma}^{\dagger} c_{\vec{i}, \sigma}
\right).
\end{eqnarray}
Here, $D_s(\beta)/2\pi^2$ corresponds to the superfluid density $n_s$
in units where the hopping $t=1$  \cite{Note}.
The above equation may be transformed to give:
\begin{equation}
\label{diffE1}
\frac{\partial}{\partial \beta}
    \left( \beta F(\Phi) -  \beta F(\Phi = 0) \right) \; = \;
                 \left(\frac{ \Phi }{\Phi_0}\right)^2
                 \frac{\partial} {\partial \beta}
                  \left( \beta D_s(\beta) \right) +
                  O\left(\left(\frac{ \Phi }{\Phi_0}\right)^4\right).
\end{equation}
The left hand side of the above equation is just the difference in
internal energy: $ E(\Phi) - E(\Phi = 0)$.
The right hand side contains the term we are interested in; namely
the temperature derivative of the superfluid density. For an infinite
lattice, the jump in the superfluid density at $T_{KT}$ implies that
\begin{equation}
        \frac{\beta \partial D_s(\beta)} { \partial \beta} \sim
         \delta \left(  T - T_{KT} \right).
\end{equation}
For a finite-size lattice, the delta-function peak scales as the number
of sites $N$ \cite{Himbergen}.

To determine which values of $\Phi$ are most appropriate for our
calculations, we consider the $T=0$ limit.
At zero temperature, the attractive Hubbard model is expected to show
off-diagonal long-range order (ODLRO), and thus flux quantization
\cite{Yang}. Hence,
the quantity
\begin{equation}
           E_0(\Phi) - E_0(\Phi=0) =
          \lim_{ \beta \rightarrow \infty}
          \lim_{ L \rightarrow \infty}
              \left( F(\Phi) -  F(\Phi = 0) \right)
\end{equation}
scales in the thermodynamic limit to a periodic function
of period $\Phi_0/2$. Furthermore, a non-vanishing energy barrier is
to be seen between the flux minima at $\Phi=0$ and $\Phi = \Phi_0/2$
\cite{Byers,Yang}. Thus, the maximal response in $E_0(\Phi) - E_0(\Phi=0)$
should lie at $ \Phi = \Phi_0/4 $.
The superfluid density, which corresponds to the
curvature of the $E_0(\Phi) - E_0(\Phi=0)$ function at $\Phi = 0$
or $\Phi=\Phi_0/2$ \cite{Yang,Scalapino}, may be approximated by:
\begin{equation}
\label{DS0}
      \frac{D_s(T=0)}{4^2}  \sim
           E_0(\Phi = \Phi_0/4 ) - E_0(\Phi = 0)
                \equiv
           E_0(\Phi = \Phi_0/4 ) - E_0(\Phi = \Phi_0/2).
\end{equation}
The above relations hold for the bose condensation of free
charge two bosons on a square lattice. Numerical evidence for  the
validity of equation (\ref{DS0}) for the quarter-filled and half-filled
two-dimensional attractive Hubbard model may be found
in reference \cite{Assaad}.

Due to the above considerations, we have computed the
quantity:
\begin{equation}
\label{diffE2}
	\Delta E(R) =  E(\Phi = \Phi_0/4 ) - E(\Phi = R) \;
\where \; R=0 \; \and \; R = \Phi_0/2
\end{equation}
as a function of temperature and lattice size.
$ \Delta E(R) $ yields at finite-temperature information
on $ \frac{ \partial \beta D_s(\beta) } {\partial \beta}$ (\ref{diffE1})
and at  $T=0$  converges  for increasing lattice size to
the superfluid density (\ref{DS0}).
Considering both $  R=0 \; \and \; R = \Phi_0/2 $
yields a cross-check for finite
size effects since  both values of $R$ converge to the same results
only in the thermodynamic limit.
As mentioned previously,  the internal energies $E(\Phi)$ were
calculated through the use of the finite temperature QMC algorithm
\cite{Hirsch,White}. We carry out a discrete Hubbard-Stratonovitch (HS)
transformation of the on-site interaction term in (\ref{Ham1}).
After integration of the fermionic degrees of freedom, one obtains
for the attractive Hubbard model:
\begin{equation}
	Z = \tr e^{-\beta H(\Phi) } =
            \sum_{\bf s} \left( \det M({\bf s}) \right)^2
\end{equation}
where $ {\bf s} $ denotes a HS configuration  and $M$ is an $L \times L$
matrix.   For $\Phi = 0$ or $\Phi = \Phi_0/2$ (i.e. antiperiodic
boundary in the x-direction), $ M $ is real and thus no
sign problem occurs. On the  other hand, for $\Phi = \Phi_0/4$,  $M$ is
complex and thus, $\left( \det M({\bf s}) \right)^2$ is equally complex.
Hence, this value of the flux yields a sign (or in our case a phase)
problem. This phase problem inhibits us of reaching very low temperatures.
Another problem we encountered, is that of fixing the chemical
potential.
Since we  are interested
in the difference of two internal energies at a given particle
number (\ref{diffE2}), the chemical potential has to be determined
very precisely. Apart from the half-filled band case where $\mu = 0$
a brute force search for the desired chemical potential is numerically
too expensive. We thus carried
out at least three simulations with chemical potential corresponding
approximately to the desired filling and fit the energy data to the
form: $ A + Bx + Cx^2$. The energy corresponding to the desired filling
is then extrapolated from the fit.  Thus, one data point
in the $\Delta E(R)$ curve corresponds to at least six independent
simulations.

Figure 1 plots $ \Delta E(R)$ for both considered values of $R$
(see equation (\ref{diffE2})), and for the half-filled attractive Hubbard model
at $U/t=-4$  as a function of the temperature. The data points at
$T=0$ were obtained with the projector QMC algorithm (PQMC) \cite{Proj}.
Here, we consider three lattice sizes: $ 4 \times 4 $, $6 \times 6 $ and
$ 8 \times 8 $. At exactly half-band filling, the attractive Hubbard
model may be mapped onto the repulsive Hubbard model by carrying out a
particle-hole transformation in one spin sector. Under this
transformation, one notices  that the s-wave pair-pair
(charge-charge) correlations of the attractive model map onto the
transverse (z-component) spin-spin correlations of the repulsive model.
In the strong-coupling limit, the repulsive Hubbard model may be
mapped onto the Heisenberg model which shows no KT transition
\cite{Ding}.  Hence, the strong-coupling attractive Hubbard model
at half-band filling
shows no KT behavior.  In the intermediate coupling regime, the
numerical data of figure 1, equally display no sign of
a KT transition. Consider
the $R=\Phi_0/2$ curves. For this value of $R$, there is a bump
appearing approximately  at $ T \sim 0.2 t$. However, this structure
fails to scale with increasing lattice size but rather appears to be
suppressed.
A similar, but less pronounced structure may be seen in the
two largest lattice sizes of the $R=0$ curves.

Let us now turn to three-quarters band filling. Our QMC data are plotted
for $R=0$ ($R=0.5$)  in figure 2a (2b). For the considered coupling
($U/t = -4 $) and band-filling ($<n>=0.75$), we were unable to reach inverse
temperatures higher than $\beta t = 7$ due to severe phase problems.
Again, the data points at $T=0$ were obtained with the PQMC algorithm
\cite{Proj}.  Figures 2a and 2b show a striking difference between the
$4 \times 4$ and $8 \times 8$ clusters.
The slope of the $\Delta E(R)$
curves become very large for the $ 8 \times 8 $ cluster size in the
temperature region $ 7 < \beta t < 5 $ \cite{Note1}.
Finally, the available data, are compatible with the onset of a Dirac
delta-function type response in $\Delta E(R)$ for both considered values of
$R$ at approximately $ T \sim 0.1t$.  This temperature value compares
favorably with the KT transition temperature obtained from a
finite-size scaling
analysis of the pair-pair correlation functions \cite{Moreo}.

In conclusion, extensions of the grand-canonical QMC
algorithm have  allowed us to
calculate the temperature
derivative of the superfluid density for the half and three-quarters
filled attractive Hubbard model at $U/t = -4$. Comparing the numerical
results for both band fillings provides convincing evidence that
the three-quarters filled attractive Hubbard model at $U/t = -4$ has a
KT transition  to a superconducting state. In contrast, no signature of
the KT transition is found at half-band filling.

We wish to thank A. Muramatsu, J.Deisz, J. Stein for instructive
conversations.
F.F. Assaad would like to thank the DFG for financial
support under
the grant number Ha 1537/6-1, W. Hanke the financial support of
the Bavarian "FORSUPRA" program on high-$T_c$ research,
and D.J. Scalapino financial support from the NSF under the grant number
DMR92-25027.
The calculations were performed on the Cray YMP of the HLRZ in
J\"{u}lich as well as on the Cray YMP of the LRZ in M\"unich.

\subsubsection*{Figure captions}
\newcounter{bean}
\begin{list}%
{Fig. \arabic{bean}}{\usecounter{bean}
                   \setlength{\rightmargin}{\leftmargin}}

\item   $E(\Phi = \Phi_0/4) - E(\Phi = R ) $ as a function of
temperature for the half-filled attractive Hubbard model at
$ U/t=-4$. Here, we consider three lattice sizes:
$4 \times 4$, $ 6 \times 6$ and $ 8 \times 8$.
The three upper (lower) curves correspond to $R=\Phi_0/2$
($R=0$).  The solid lines are a spline fits to the data. The data
points at $T=0$ were obtained with the PQMC algorithm.

\item   a) $E(\Phi = \Phi_0/4) - E(\Phi = 0 ) $ as a function of
temperature for
the three-quarters filled attractive Hubbard model at $U/t = -4 $.
The data points at $T=0$ were obtained with the PQMC algorithm.
The solid lines are spline fits to the data.\\
b) Same as figure a) but we consider:
$E(\Phi = \Phi_0/4) - E(\Phi = \Phi_0/2) $.

\end{list}


\begin{thebibliography}{99}
\bibitem{Note}  Note that to obtain the  $T=0$ superfluid density
from the expression for $D_s(\beta)$ in equation (\ref{Expan}) one
should first take the limit $L \rightarrow \infty $ and then
the limit $ \beta \rightarrow \infty $. This ordering of the limits
removes the ambiguity between the superfluid density and the
Drude weight. (see also ref. \cite{Scalapino})
\end{thebibliography}
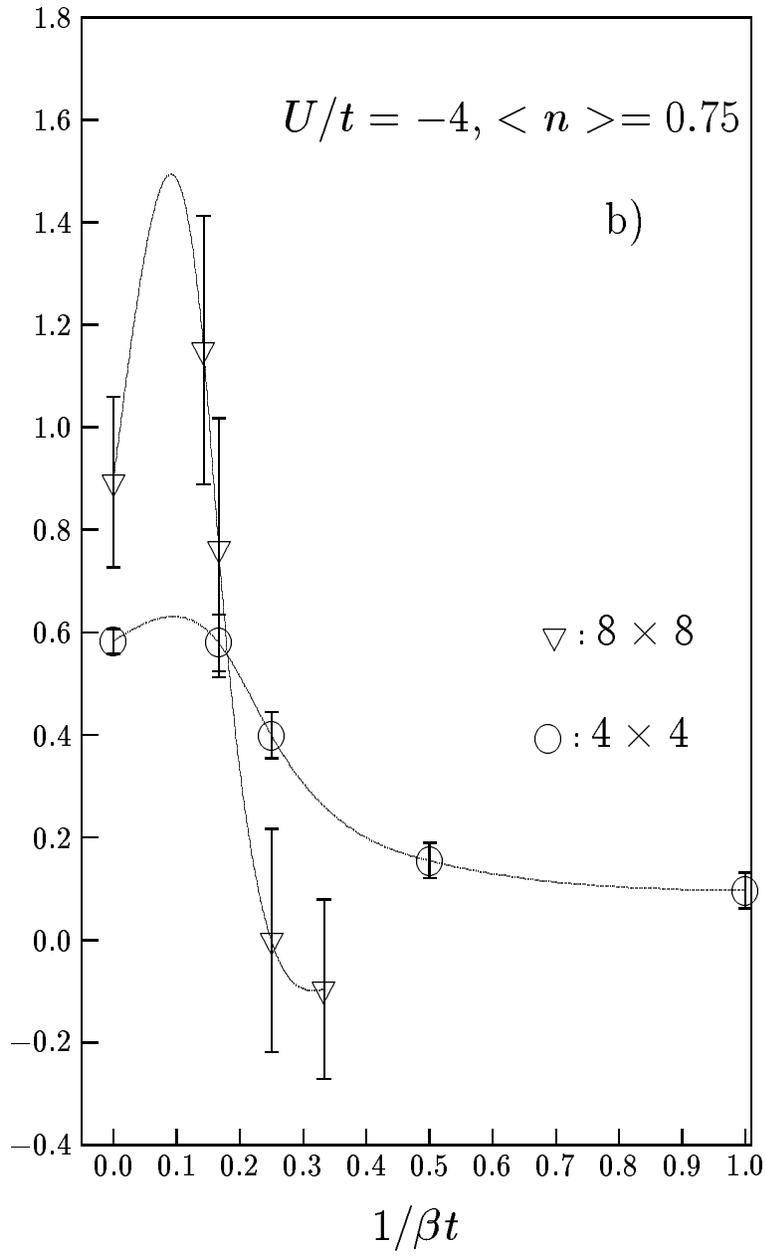
\end{document}